\UseRawInputEncoding
\documentclass[twocolumn,preprintnumbers,amsmath,amssymb]{revtex4-1}
\usepackage{epsfig}
\usepackage{color}
\usepackage{amsmath}
\usepackage{multirow}

\begin{document}

\title{Equivalence between condensation and boiling in a Lennard Jones fluid}

\author{I. Sanchez-Burgos$^1$, P. Montero de Hijes$^1$, P. Rosales-Pelaez$^1$, C. Vega$^1$ and E. Sanz$^1$}
\affiliation{$^1$Departamento de Qu\'{\i}mica F\'{\i}sica,
Facultad de Ciencias Qu\'{\i}micas, Universidad Complutense de Madrid,
28040 Madrid, Spain}

\begin{abstract}

Condensation and boiling are phase 
transitions highly relevant to industry, geology  
or atmospheric science. 
These phase transitions are initiated by the nucleation of a drop in a supersaturated
vapor and of a bubble in an overstretched liquid respectively. The surface tension 
between both phases, liquid and vapor, is 
a key parameter in the development 
of such nucleation stage. 
Whereas the surface tension can be readily  measured
for a flat interface, there are technical and 
conceptual limitations to obtain it for the 
curved interface of the nucleus. On the technical
side, 
it is quite difficult to observe a critical 
nucleus in experiments. From a conceptual point of view, the
interfacial free energy 
depends on the choice of the dividing surface, being the surface of tension the one  relevant for nucleation.
We bypass the technical limitation by performing
simulations of a Lennard Jones fluid where
we equilibrate critical nuclei (both drops and 
bubbles). Regarding the conceptual hurdle, we
find the relevant cluster size by searching the
radius that correctly predicts nucleation 
rates and nucleation free energy barriers
when combined with Classical
Nucleation Theory. 
With such definition of the cluster size we
find the same value of the surface tension 
for drops and bubbles of a given radius.
Thus, condensation and boiling can be viewed as
two sides of the same coin.
Finally, we combine the data coming from drops and bubbles
to obtain, via two different routes, 
estimates of the 
Tolman length, a parameter that allows
describing the curvature dependence of the
surface tension in a theoretical framework. 

\end{abstract}

\maketitle

\section{Introduction}

Understanding first order phase transitions is of great importance
to many fields, ranging from biology \cite{cryopres}, to atmospheric science \cite{cantrell2005production}, physics \cite{debenedettibook}, geology \cite{icemars2018} or industry \cite{foodcrystallization,brittain}. 

In the absence of impurities or external surfaces, 
first order phase transitions start with the
emergence of a nucleus of the stable phase in the bulk 
of the parent metastable phase \cite{kelton,skripov1974metastable}.
A nucleus is ``critical''  if it is big enough 
so that it has 50 per
cent chances to either grow or redissolve.

Although the emerging phase is more stable, the
presence of an unfavourable interface between the 
nucleus and the 
parent phase can 
delay to a great extent the phase transition.
Thus, for instance, alkane vapors can be saturated
thousands of times over their
vapor pressure before condensation takes place \cite{wyslouzil2016overview}, alkane liquids
can be substantially superheated above the boiling temperature \cite{skripov2010phenomenon,lipnyagov2018going,skripov1992metastable}
or liquid water can be supercooled up to $\sim 60$
K below melting until it freezes \cite{manka2012,amaya2018ice,hu2008water,bhabheJPCA2013}.

Therefore, the surface tension, $\gamma$, or the free energy per unit
area between both phases, plays a key role 
in the development of first order phase transitions. 
Whereas $\gamma$ can be readily measured for a flat
interface at equilibrium --at least between fluid phases \cite{ickespccp2015}-- 
it cannot be directly probed for curved interfaces, which
is the relevant case for nucleation.
Moreover, the fact that critical nuclei are nanoscopic objects
makes it very difficult to observe them in experiments, 
let alone measuring their $\gamma$.
The usual strategy is to infer $\gamma$ 
by combining a theoretical description of nucleation
with measurements of the 
nucleation rate
(the number of nuclei that appear per unit of time and volume)
\cite{kelton,ickespccp2015,caupinPRL2016curvature,granasy:6157}. This approach relies on the validity of 
theoretical approximations that are difficult to assess. 

Computer simulations do have access to the time and
length scales relevant for the observation of critical
nuclei. However, 
whereas the methodology and theoretical 
framework for computing $\gamma$ for 
flat interfaces is very well established
\cite{kirkwood1949statistical,JCP_2005_123_134703,broughton:5759,PhysRevLett.86.5530,PNAS_2002_99_12562,ubertiPRB2010,verrocchioPRL2012,PhysRevLett.94.176105,Nature_2001_409_1020,espinosaJCP2014_2}, 
that for curved interfaces is still under development \cite{lau2015surface,ceriottitolman2018,binder2012beyond}.
One of the key issues is that $\gamma$ for 
curved interfaces depends on the location of the
interface, that can be defined in different ways \cite{rowlinson2013molecular,binder2012beyond}.
The current situation is that the dependence of $\gamma$
with the curvature of the interface is contradictory 
between different groups \cite{blokhuis2006thermodynamic,block2010curvature,sampayo,malijevsky2012perspective,binder2012beyond,size-dependendentgamma2005,ceriottitolman2018,JCP_1984_81_00530,lau2015surface,vrabec2006comprehensive,wilhelmsen2015tolman,joswiak2016energetic,schmelzer2019entropy,richard2018crystallization,caupinPRL2016curvature}

In this work we address fundamental questions
regarding the liquid-vapor interface with computer
simulations.
It has been shown in different simulation works that spherical nuclei
can be equilibrated at constant volume and temperature in 
finite systems \cite{block2010curvature,schrader2009simulation,matsumoto2008nano,troster2012numerical,macdowell2006nucleation,richard2018crystallization,statt2015finite,koss2018phase,gunawardana2018theoretical,zierenberg2015exploring,zierenberg2017canonical}. Recently, we showed with simulations of bubbles \cite{seedingNVT}
and crystals \cite{montero2020interfacial}
that nuclei thus equilibrated are critical, in agreement with Density Functional Theory (DFT) predictions \cite{lutsko2018classical,lutsko2019crystals}. 
On the other hand, we have extensively developed
in the past years the so-called Seeding method
\cite{jacs2013,knottJACS2012,baiJCP2006,seedingvienes}
to study nucleation phenomena. This method consists
in obtaining with simulations the properties of 
critical clusters and ``plug'' them in
the Classical Nucleation Theory (CNT) formalism \cite{ZPC_1926_119_277_nolotengo,becker-doring,gibbsCNT1,gibbsCNT2}
to obtain predictions of the nucleaiton 
rate and of the $\gamma$ curvature dependence. 
This approach has been successful for a wide range
of systems \cite{knottJACS2012,jacs2013,seedingvienes,zaragozaJCP2015,espinosaPRL2016,espinosaJPCL2017,espinosa2018homogeneous,seedingNpT,seedingNVT} and we use it here for the 
first time to study condensation. In particular, we apply 
Seeding at constant volume both to condensation and to
cavitation for a Lennard Jones model.

Since Seeding relies on CNT, it is necessary to 
validate it by 
comparing its predictions with  
independent calculations that do not rely on such 
framework. We do so by computing 
nucleation rates via Umbrella Sampling (US) \cite{torrie1974monte,JCP_1992_96_4655} and
direct brute force simulations \cite{filion:244115} as well as by 
testing the consistency of the $\gamma$-curvature
dependence obtained via Seeding with the
value for a flat interface. 

All consistency tests are successfully passed for 
our Seeding simulations provided
that the nucleus surface is identified with that 
where 
the density is the average between the density of both phases
(``equi-density'' surface). 
Therefore, we identify the equidensity surface with the surface of tension.
On the other hand, we directly compare the condensation of
liquid drops in a supersaturated vapor with 
the cavitation of vapor bubbles in an overstretched liquid.
We find that, for a given 
temperature, drops and bubbles of
the same radius have the same $\gamma$ when using the equidensity definition
of the surface of tension.
Finally, we estimate the Tolman lenght \cite{tolman1949effect},
a parameter useful to predict the $\gamma$-curvature dependence, via two different routes, as recently proposed in
\cite{montero2020interfacial}.

\section{Simulation details}
The Lennard Jones model potential, as well as 
the simulation details, are the same as  
in our previous work \cite{seedingNpT,seedingNVT}. 
In particular, we study
the truncated and force-shifted Lennard-Jones
(TSF-LJ) potential \cite{wang2008homogeneous}, a model for which
the vapor-liquid transition has been previously investigated \cite{wang2008homogeneous,tanaka2015simple,meadley2012thermodynamics,seedingNpT}:
\begin{equation}
U_{TSF-LJ}(r) = U_{LJ}(r) - U_{LJ}(r_{c}) - (r-r_{c})U'_{LJ}(r_{c}), 
\end{equation}
where $U_{LJ}(r)$ is the standard 12-6 Lennard-Jones potential and $U'_{LJ}(r)$ is its
first derivative. The interaction potential is truncated and shifted at  $r_{c}=2.5 \sigma$,
being $\sigma$ the particle's diameter  and $\epsilon$ the depth of the un-truncated Lennard-Jones potential.
Unless otherwise specified, 
all magnitudes in this work are given in
Lennard-Jones reduced units \cite{seedingNVT}.
Thus, the reported temperatures are reduced
by $\epsilon/k_B$, distances by $\sigma$, densities
by $\sigma^{-3}$, pressures by $\epsilon/\sigma^3$, times
by $\tau=\sqrt{m \sigma^2/\epsilon}$ (being $m$ 
the particle mass), interfacial free 
energies by $\epsilon/\sigma^2$ and nucleation 
rates by $1/(\tau \sigma^3)$.

We use cubic boxes with periodic boundary condition
and the Molecular Dynamics (MD) LAMMPS package \cite{lammps_program} to perform all simulations of
this work.
The equations of motion are integrated with a 
leap-frog algorithm \cite{leapfrog}.

In the MD Seeding simulations we used 
a time-step of $0.0012$.
The system was kept at constant temperature 
using the Nos\'e-Hover thermostat \cite{JCP_1984_81_00511}
with a relaxation time of 0.46.

For the MD  simulations used within the Umbrella Sampling
scheme we set the 
time step for the integration of the motion equations to $0.0012$.
The relaxation times for the Nose-Hover thermostat
and barostat were 0.46 and 1.6 respectively. 

All simulations are carried out at T=0.785. 
The coexistence pressure at such temperature 
for the model is $p_{coex}$=0.0267. We determined
this value, refined with respect to that of 
0.026 previously published \cite{wang2008homogeneous},
by running long ($4 \cdot 10^5$ $\tau$) MD NVT simulations with an elongated
box (50 x 17 x 17) where the vapor and the liquid were put at
contact at the temperature of interest. The average
pressure normal to the interface in such 
simulation corresponds to $p_{coex}$.

\section{Seeding of condensation}
\label{cnt}

This work is based on a recent publication
where we demonstrate how to compute 
bubble nucleation rates in an overstretched 
Lennard-Jones fluid by equilibrating 
critical bubbles in the NVT ensemble, an approach we call ``NVT-Seeding'' \cite{seedingNVT}.
The Seeding method, originally developed to 
study crystal nucleation \cite{jacs2013,knottJACS2012,baiJCP2006,seedingvienes},
and more recently applied to vapor cavitation \cite{seedingNpT,seedingNVT,baidakov2020molecular}, consists in 
combining CNT \cite{ZPC_1926_119_277_nolotengo,becker-doring,gibbsCNT1,gibbsCNT2}
with computer simulations to estimate nucleation free energy barrier heights, 
$\Delta G_c$, interfacial free energies, $\gamma$, and, most importantly, nucleation rates, $J$. 

According to CNT, the Gibbs free energy barrier for
the nucleation of a spherical liquid drop is given by
the following expression:
\begin{equation}
\Delta G = \gamma A - V \Delta p
\label{cnteq}
\end{equation}
Where $V$ and $A$ are the volume and the area of the drop 
respectively. 
By maximizing 
Eq. \ref{cnteq} assuming a spherical drop shape one 
obtains both the height of the nucleation free energy barrier, 
\begin{equation}
\label{eq:deltag}
    \Delta G_c = \frac{2 \pi R_c^3 \Delta p}{3},
\end{equation}
where $R_c$ is the critical droplet radius and $\Delta p$ 
is the pressure difference between the interior of the 
drop and the surrounding vapor, and 
the number of particles in the critical drop,
\begin{equation}
N_c=(32 \pi \rho_l \gamma^3)/(3 \Delta p^3),
\label{eqnc}
\end{equation}
where $\rho_l$ is the critical drop number density and
$\gamma$ is the liquid-vapor surface tension.
By substituting in the equation above $N_c$ by 
the droplet volume ($4/3\pi R_c^3$) times $\rho_l$
one recovers the Laplace equation:
\begin{equation}
    \Delta p = \frac{2 \gamma}{R_c}.
    \label{laplaceeq}
\end{equation}
This derivation shows that the Laplace equation, which 
is valid when the droplet surface is located at the
the surface of tension, is implicit in CNT. Consequently, 
$R_c$ should be identified with the radius of tension, 
$R_s$. This is an important point that we will use later
on in the paper. 

The CNT prediction for the nucleation rate of drops is given by \cite{kelton}: 
\begin{equation}
J= A_0 \rho_{vap} \exp \left (-\frac{\Delta G_c} {k_B T} \right ), 
\label{eq:rate}
\end{equation}
where $k_B$ is the Boltzmann constant, $\rho_{vap}$ is the density of the vapor phase that multiplied by $\exp \left (-\frac{\Delta G_c} {k_B T} \right )$
gives the number density of critical clusters and 
$A_0$ is a kinetic pre-factor.

$A_0$ is computed as
the product of the Zeldovich factor, $Z$, and 
the rate of attachment to the critical nucleus, $f^+$ \cite{becker-doring,kelton}:
\begin{equation}
A_0= Z f^+.
\label{eqA0}
\end{equation}

$Z$ takes into account the establishment of
a steady state and, 
according to CNT, is given by  \cite{kelton,ZPC_1926_119_277_nolotengo,becker-doring}:
\begin{equation}
    Z=\sqrt{\frac{|\Delta G(N)^{''}|_{N_c}}{2 \pi k_B T}}=\sqrt{\frac{\Delta p}{6 \pi k_B T \rho_l N_c}}=\sqrt{\frac{\Delta p}{8 \pi^{2} k_{B}T \rho_{l}^{2} \cdot R_c^{3}}}
    \label{zeq}
\end{equation}
where $N_c$ is the number of particles in the drop  and $|\Delta G_c(N)^{''}|_{N_c}$ is the curvature of $\Delta G(N)$ evaluated at the barrier top.

The attachment rate, $f^+$, can be estimated by multiplying
the collision frequency of the vapor per unit of wall area 
given by the 
kinetic theory of gasses (ktg) by 
the area of the critical bubble:
\begin{equation}
    f^+_{ktg}=\sqrt{\frac{k_BT}{2\pi m}} \left(\frac{6\sqrt{\pi} N_c}{\rho_l}\right)^{2/3}
    \label{fktgeq}
\end{equation}
where the subscript
``$ktg$'' stresses the fact that this expression
of the attachment
rate is based on the kinetic theory of gasses. 

Combining this equation with \ref{zeq}, \ref{eqnc},
and \ref{laplaceeq},
the following kinetic pre-factor is obtained:
\begin{equation}
    A^{ktg}_{0}=\sqrt{\frac{\Delta p R_{c}}{\pi m}} \frac{\rho_{vap}}{\rho_l}.
    \label{eq:a0ktg}
\end{equation}

\begin{figure}[h!]
    \centering
    \includegraphics[width=0.85\linewidth]{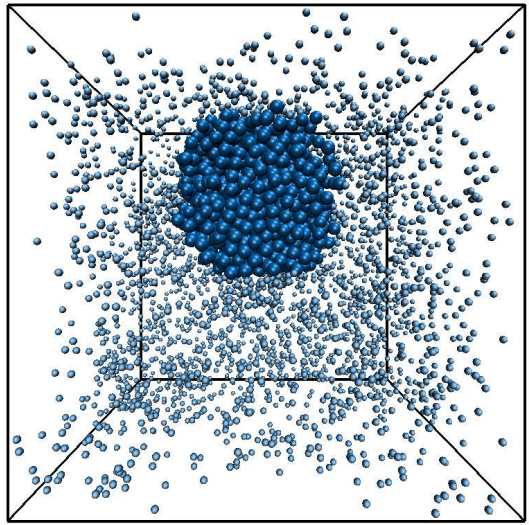}
    \caption{Snapshot of a critical drop equilibrated in the NVT ensemble at T=0.785 surrounded by supersaturated vapor. The droplet radius is about 6.8 and the density of the surrounding vapor 0.0550.}
    \label{gota}
\end{figure}

The equations above are quite powerful, because only 
$R_c$, $\Delta p$ and the density of
both phases are required
to obtain key nucleation parameters as free energy barriers, interfacial free energies and \i
nucleation rates. The Seeding method
consists in performing simulations
of a cluster of the stable phase surrounded by the mestastable
phase (a liquid drop surrounded by supersaturated
vapor in our case as shown in Fig. \ref{gota})
to compute $R_c$, $\Delta p$, $\rho_l$ and $\rho_{vap}$ 
in order to get 
``cheap'' estimates of $\Delta G_c$, $\gamma$ and, 
most importantly, $J$ through the expressions above. 

\textcolor{black}{The main drawback of Seeding is that the definition of $R_c$ is not unique. Therefore, the resulting free energy barrier
depends on the specific definition of $R_c$. This contrasts with rigorous simulation methods like Umbrella Sampling \cite{JCP_2005_122_194501,filion:244115}
    or with theoretical approaches like DFT \cite{shen2001density,gallo2020nucleation,lutsko2008density} where the nucleation free energy
does not depend on the criterion chosen to measure the nucleus size, which can be estimated \emph{a posteriori} via, e. g. the nucleation theorem \cite{JCP_1982_76_05098,kashchiev2006forms,wedekind2007best}
(although in the particular case of DFT an approximate functional needs to be proposed so that the results do also contain approximations). To 
assess the suitability of our choice to compute $R_c$ we 
complement Seeding with Umbrella Sampling simulations.}

\subsection{$R_c$, $\Delta p$, $\rho_l$ and $\rho_{vap}$}
\label{sec:rc}

We use the $NVT$ ensemble to run the simulations
of the drops given that in such ensemble  
critical nuclei are naturally equilibrated and 
stabilised for long times \cite{seedingNVT,montero2020interfacial}.
We equilibrate drops in 10 different 
systems. The edge of the cubic simulation 
box, $L$, and the total number of 
particles in each system, $N_T$, are 
reported in table \ref{table:sizes}.
A large number of particles is used to minimize
finite size effects \cite{zierenberg2015exploring,marchio2018pressure,seedingNVT}.
Each system was simulated for about  $10^3$ 
Lennard Jones times 
of equilibration and $2\cdot 10^5$ of production. 

To prepare the initial configuration we cut a
spherical liquid drop  from 
a bulk liquid simulation and insert it in a bulk
vapor box removing the overlapping
vapor particles. 
The liquid drop is cut with a certain tentative 
radius, but
the precise number of particles in each 
phase 
is not crucial given that 
equilibrium is reached along the course of the
$NVT$ simulation. 

\begin{figure}[h!]
    \centering
    \includegraphics[width=0.85\linewidth]{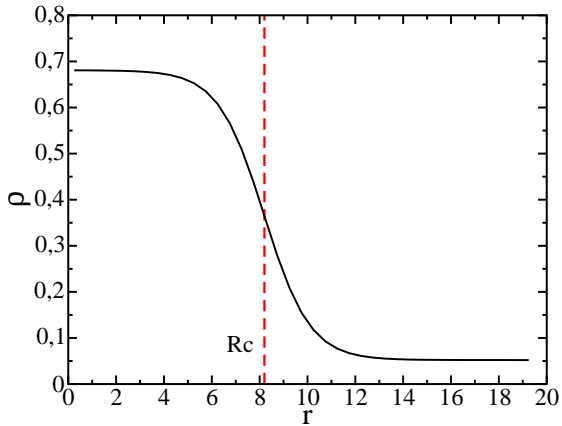}
    \caption{Density profile of a critical drop equilibrated in the NVT ensemble at T=0.785 surrounded by supersaturated vapor. The droplet radius, indicated by a red 
     vertical line in the figure, is given by the point
     at which $\rho(r)$ takes an average value between both plateaux (equi-density criterion). The density profile corresponds to the system labelled
     as IV in table \ref{table:sizes}}.
    \label{rdp}
\end{figure}

From a simulation of a drop surrounded
by supersaturated vapor one can obtain an average
radial density profile starting from the 
center of the drop as that shown in 
Fig. \ref{rdp} (to find the drop centre
in each configuration we use a similar strategy
to that described in our previous work \cite{seedingNVT}
consisting in this case in identifying 
the maxima of density profiles computed along each cartesian coordinate).
Following Refs. \cite{seedingNpT,seedingNVT},
we obtain $R_c$
from such density profile as the distance at which the density is average
between the liquid and the vapor plateaux. This is 
indicated with a vertical dashed line 
in Fig. \ref{rdp}. We refer to this way
of obtaining $R_c$ as the ``equi-density'' criterion. 
The $R_c$'s thus obtained in our NVT-Seeding
simulations are also reported in Table \ref{table:sizes}.
Other definitions of $R_c$
are in principle \textcolor{black}{as valid
as the equi-density criterion \cite{seedingNpT,lutsko2008density,gallo2020nucleation}. We
argue later on in the paper that our $R_c$ definition
is} a good one
because it makes Seeding predictions 
consistent with independent calculations
of $\gamma$, $J$ or $\Delta G_c$.

\begin{figure}[h!]
\includegraphics[width=0.9\linewidth]{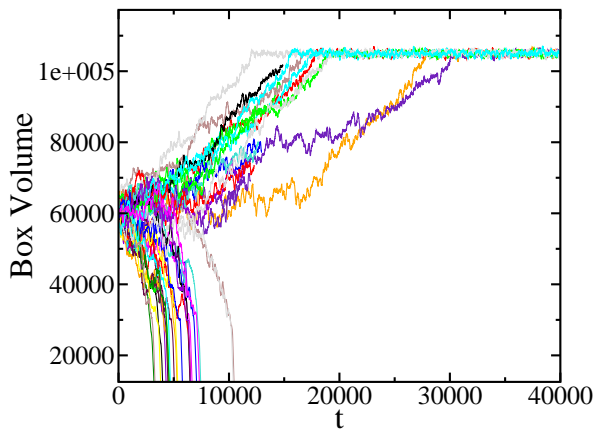}
\caption{\textcolor{black}{Box volume versus time in NpT simulations starting from 40 configurations taken from the NVT-Seeding simulation labelled as IX in Table \ref{table:sizes}. The imposed pressure is
the average viral pressure of the NVT-Seeding run.}}
\label{fig:npt}
\end{figure}

\begin{figure}[h!]
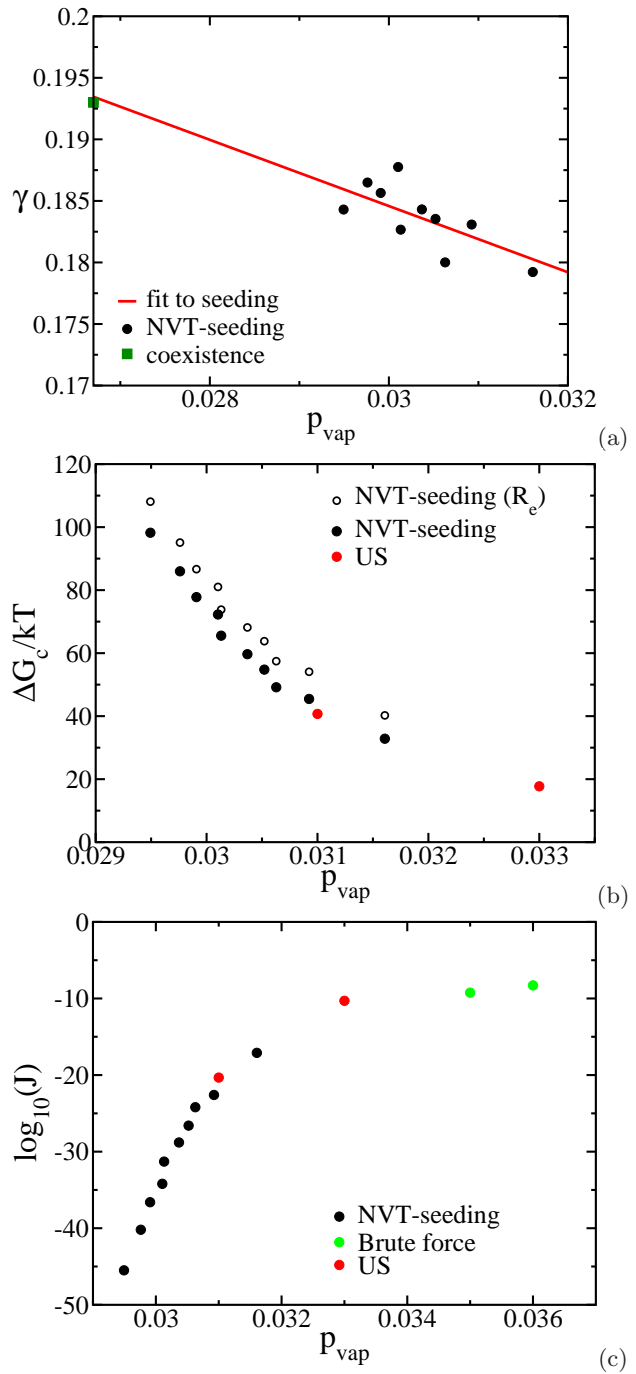

\includegraphics[width=0.9\linewidth]{gamma_pvap.eps}(a)\\
	\includegraphics[width=0.9\linewidth]{dG_p.eps}(b)\\
		\includegraphics[width=0.9\linewidth]{J_pvap.eps}(c)
	\caption{(a) $\gamma$ vs vapor pressure obtained from NVT-Seeding data of droplets surrounded by supersaturated vapor. The surface tension at coexistence (p=0.0267) is included \cite{seedingNpT}.
	(b) $\Delta G_c$ vs vapor pressure. NVT-Seeding and US data are compared. 
	Empty black symbols correspond to Seeding predictions when
	the Gibbs dividing (equi-molar) --instead of the equi-density-- surface is employed
    to identify the cluster radius. 
	(c) Nucleation rate versus vapor pressure as obtained from NVT-Seeding, US and spontaneous nucleation.}
\label{deltaG}
\end{figure}

To get  $\Delta p=p_l - p_{vap}$ we
obtain first the vapor density, $\rho_{vap}$, 
by counting the number of particles outside a sphere concentric 
with the drop but with a larger 
radius (we use a sphere radius 7$\sigma$ larger than that of the drop, 
but
we have checked for a few selected cases 
that any value beyond $\sim 5 \sigma$ gives
the same result). $\rho_{vap}$ is given by 
the number of particles outside
the sphere divided by the $L^3$ minus
the sphere volume.
We then use the bulk vapor equation of state to infer $p_{vap}$ from
$\rho_{vap}$.
We report $p_{vap}$ and $\rho_{vap}$ in table \ref{table:sizes}. 
We have checked for all studied systems that $p_{vap}$ coincides
with the overall virial pressure of the system.
On the other hand, $p_l$
is obtained, as in our previous work \cite{seedingNpT,seedingNVT}, by assuming equal chemical 
potential between the critical drop and the surrounding
vapor:
\begin{equation}
\int^{p_{vap}}_{p_{coex}} \frac{1}{\rho_{vap}(p)} dp =
\int^{p_{l}}_{p_{coex}} \frac{1}{\rho_{l}(p)} dp
\label{eq:intterm}
\end{equation}
where $p_{coex}$ is the coexistence pressure and
$\rho_{vap}(p)$ and $\rho_{l}(p)$ are the
bulk vapor and bulk liquid number densities 
at pressure $p$. In table \ref{table:sizes}
we report $p_l$ and $\Delta p$  for 
all studied systems. Once $p_l$ is known, 
$\rho_l$, also reported in the table, can be easily computed from the 
bulk liquid equation of state. In all cases, this computation of $\rho_l$,
based on the equality of chemical potential between both phases,
is consistent with that obtained from the density profiles. For instance, 
for system IV we get $\rho_l = 0.0680$, which is fully consistent
with the first plateau observed in the density profile shown in 
Fig. \ref{rdp}. This means that the mechanical pressure and the thermodynamic
pressure inside the drop coincide, a matter of current debate for solid-liquid
nucleation \cite{presiontermod}.

\textcolor{black}{There has
been much 
simulation, theoretical and experimental work 
devoted to study the formation of 
nuclei confined at constant volume \cite{vincent2017statics,marti2012effect,macdowell2006nucleation,finitesizenucreguera,reguera2004fusion,JCP_2003_118_00340,binder2012beyond,bindernuccryst2017,richard2018crystallization,yang1985thermodynamical,gallo2018thermally,gallo2020heterogeneous}.
In Refs. \cite{seedingNVT,montero2020interfacial} we showed with simulations that
nuclei equilibrated in the NVT ensemble are critical 
because they
have equal chances to grow or shrink when simulated in the NpT ensemble at the same temperature and at
the average pressure along the NVT run.
Based on this result, we opted to study here 
drop nucleation in the NVT ensemble, 
where statistics is better because clusters remain stable for very long times \cite{seedingNVT}. 
Stabilising nuclei to gain time to study their
properties is something quite desirable. 
An alternative strategy to the use of 
constant volume simulations is to pin
the nucleus to a heterogeneous solid 
substrate \cite{xiao2017experiments}.}

\textcolor{black}{
Despite having already shown the 
equivalence between stable (NVT) and critical (NpT) nuclei for cavitation \cite{seedingNVT}, we check here for one of the NVT-Seeding simulations
if the drops equilibrated at constant volume and
temperature do correspond to a Gibbs free energy maximum.
In Fig. \ref{fig:npt} we show the evolution of the 
box volume in NpT simulations started from 40 independent
configurations gathered along the NVT-Seeding trajectory
labelled as IX in Table \ref{table:sizes}.
The imposed pressure is the average virial pressure along
the NVT-Seeding run. 
Roughly, 
in 50 per cent of the cases the box expands (the drop dissolves) and in the other half of the cases the box 
shrinks (the drop grows). This result supports the use
of NVT to study drop condensation in the same manner we did
for bubble cavitation and cystal nucleation \cite{seedingNVT,montero2020interfacial}. 
Furthermore, the equivalence between clusters
equilibrated at constant volume and critical
nuclei has been recently proven 
with DFT theoretical 
arguments for crystallization (see supplementary material of Ref. \cite{lutsko2019crystals}}).

\begin{table*}[]
\begin{ruledtabular}
\begin{tabular}{cccccccccccccccc}
Label &
$L$ &
$N_T$ &
$\rho_{l}$  &
$\rho_{vap} $ &
$p_{l}$ &
$p_{vap}$ &
$\Delta p$ &
$R_c$ &
$\gamma$ &
$\Delta G_c/(k_BT)$ &
$A_{0}^{ktg}$  &
$A_{0}^{af}$  &
$\log_{10}(J)$\\

\hline
I & 38.019 & 3774 & 0.6833 & 0.05690 & 
0.09279 & 0.03161 
& 0.0612 & 5.86 & 0.1792 & 32.8 & 0.028 & &-17.1 \\
II & 39.160 & 4291 & 0.6811 & 0.05496 & 
0.0846 & 0.03092 
& 0.0537 & 6.82 & 0.1831 & 45.5 &0.028 & &-22.6\\
III & 39.160 & 4373 & 0.6801 & 0.05414 & 
0.08095 & 0.03063
& 0.0503 & 7.15 & 0.1800 & 49.2 & 0.027 & &-24.2\\
IV & 39.160 & 4510 & 0.6797 & 0.05384 & 
0.07961 & 0.03052
& 0.0491 & 7.48 & 0.1835 & 54.8 &0.027 & 0.030 &-26.6\\
V & 39.160 & 4623 & 0.6792 & 0.05342 & 
0.07769 & 0.03037
& 0.0473 & 7.79 & 0.1843 & 59.7 &0.027 &  &-28.8\\
VI & 39.160 & 4796 & 0.6782 & 0.05277 & 
0.07468 & 0.03013 
& 0.0446 & 8.20 & 0.1827 & 65.5 & 0.027 & 0.036 & -31.3 \\
VII & 39.160 & 4964 & 0.6783 & 0.05269 & 
0.07432 & 0.03010 
& 0.0442 & 8.49 & 0.1878 & 72.3 & 0.027 & &-34.2\\
VIII & 39.160 & 5163 & 0.6776 & 0.05216 & 
0.07181 & 0.02991
& 0.0419 & 8.86 & 0.1856 & 77.8 & 0.027 & &-36.6\\
IX & 39.160 & 5435 & 0.6771 & 0.05176 & 
0.06989 & 0.02976 
& 0.0401 & 9.30 & 0.1865 & 86.0 & 0.026 &  &-40.2\\
X & 85.264 & 34519 & 0.6761 & 0.05105 & 
0.06638 & 0.02949 
& 0.0369 & 9.99 & 0.1843 & 98.2 & 0.026 &  &-45.5\\
\end{tabular}
\end{ruledtabular}
\caption{NVT-Seeding data for the different drops studied in this work at T=0.785.}
\label{table:sizes}
\end{table*}

Having computed $R_c$, $\Delta p$ and $\rho_{vap}$
and $\rho_l$
we have everything needed to obtain 
$\gamma$, $\Delta G_c$ and $J$
according to the equations presented 
in section \ref{cnt}. We report the values
for these variables in table \ref{table:sizes}
and plot them in fig. \ref{deltaG}(a)-(c) versus the vapor 
pressure with black dots. In the following section
we comment each of these graphs.

\subsection{$\gamma$, $\Delta G_c$ and $J$}

\subsubsection{$\gamma$}
As shown in Fig. \ref{deltaG}
(a) the prediction we obtain from 
Seeding is that $\gamma$ decreases as 
the vapor supersaturation increases.
This trend is in agreement with previous 
work \cite{block2010curvature,troster2012numerical}.
Accordingly, using the capillarity approximation
(i. e., that $\gamma$ is pressure independent) 
would be erroneous. 
The green square in Fig. \ref{deltaG}(a) corresponds
to the surface tension at coexistence \cite{seedingNpT} obtained through 
the pressure tensor \cite{walton1983pressure} in an NVT simulation of a liquid and a vapor at contact.
The trend of the Seeding data is fully consistent with the coexistence 
value, as shown by the linear fit in the figure.
This is a good consistency test, although 
the $\gamma$ values provided by Seeding could still be incorrect
despite the fact that they extrapolate correctly to 
coexistence. Therefore, a test for Seeding predictions 
away from coexistence
is needed. 

\subsubsection{$\Delta G_c$}
To further test our Seeding results
we compare $\Delta G_c$
obtained with Seeding with that computed
via Umbrella Sampling.
In Fig. \ref{deltaG}(b), where we plot
$\Delta G_c$ versus the vapor pressure, 
black solid dots correspond to Seeding 
and red ones to US (details
on US calculations are described in section
\ref{sec:us}). 
Whereas Seeding predictions rely on the validity of 
CNT and on a proper definition of $R_c$,  
Umbrella Sampling calculations are rigorous and 
independent on the criterion to identify 
the nucleus size \cite{filion:244115}.
On the other hand Seeding is much ``cheaper'' 
than US from a computational point of view.
As a matter of fact, Seeding has access to much 
higher nucleation barriers than US. 
The accordance between Seeding and US shown in 
Fig. \ref{deltaG} is excellent, which gives
us great confidence on Seeding predictions.
The choice of the equi-density surface to identify
the drop radius has proven correct.
If we use another criterion, such as 
the Gibbs (equi-molar) dividing surface, the agreement between 
Seeding and Umbrella Sampling deteriorates (empty black symbols in 
Fig. \ref{deltaG} (b)). To compute $R_e$, the radius
associated to the Gibbs dividing surface, we use
$N_T=N_l+N_{vap}$ where $N_l=4/3 \pi R_e^3 \rho_l$
and $N_{vap}=[V_T-(4/3 \pi R_e^3)] \rho_{vap}$, 
where $V_T$ is the volume of the simulation box and
the densities $\rho_l$ and $\rho_{vap}$ are obtained
as described in Sec. \ref{cnt}.

In a recent
publication on cavitation (nucleation of bubbles instead of drops) we compared the performance of different
criteria to identify the cluster radius and found
that the equi-density criterion also
made Seeding predictions consistent with other
rigorous calculations \cite{seedingNpT}.
Therefore, 
identifying the critical drop radius 
with the equi-density distance seems to be quite general for condensation-evaporation
transitions.

\subsubsection{J}
\label{sec:J}

Once $\Delta G_c$ is known computing
$J$ via Eq. \ref{eq:rate} is quite straight forward. 
The kinetic pre-factor $A_0$ given by 
the kinetic theory of gases, Eq. \ref{eq:a0ktg},
depends on parameters we already have under control: $\Delta p$, $R_c$
and the density of both phases. 
The values of $A_0$ computed via Eq. \ref{eq:a0ktg}, $A_0^{ktg}$, are reported in Table \ref{table:sizes}. 

These $A_0$ values are approximate since they rely on 
the validity of the kinetic theory of gasses
to estimate the attachment rate (see section \ref{cnt}).
We therefore have to check $A_0^{ktg}$
by computing the attachment rate with an alternative approach. 
Following the
work by Auer and Frenkel \cite{auerJCP2004}, the attachment rate can be computed 
from the diffusion of $N$, the number of particles in 
the liquid drop, around the critical drop \cite{auerJCP2004}:
\begin{equation}
    f^+_{af}=\frac{\left<(N(t)-N(0))^2\right>_{Nc}}{2t}
    \label{eqf+},
\end{equation}
where the average is performed over several trajectories 
starting from a critical drop configuration.
The $af$ subscript
stresses the fact that this expression of the 
attachment rate is based on the work by Auer and Frenkel.

\begin{figure}[h!]
    \centering
    \includegraphics[width=\linewidth]{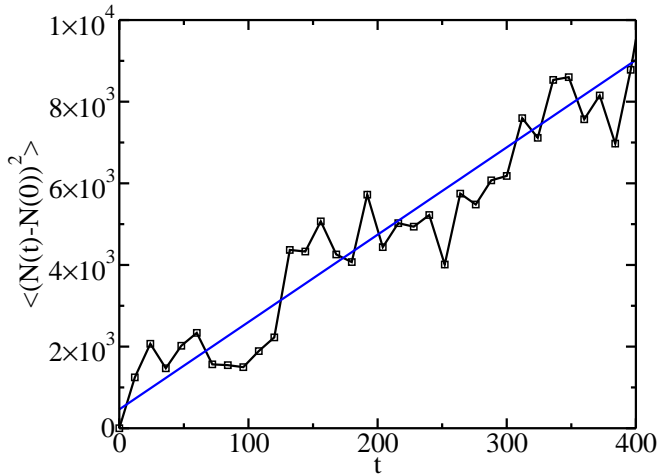}
    \caption{Time dependence of the mean squared deviation of the number of particles in the critical drop for system VI in table \ref{table:sizes}. Half the slope of this plot gives the attachment rate according to Eq. \ref{eqf+}}.
    \label{f+}
\end{figure}

To compute $N$ we follow \cite{JCP_1998_109_09901}.
We count as neighbors all 
particles within a 1.625 distance of a tagged particle. Particles
with 8 or more neighbors are labelled as ``liquid''. 
Two liquid particles belong to the same drop 
if their mutual distance is less than 1.625. 
An example of the calculation of $f^+$ according
to equation \ref{eqf+} is illustrated in Fig.  
\ref{f+}. Typically, $\left<N(t)-N(0)\right>_{N_c}$ is obtained
by averaging 20 NpT runs started 
from independent configurations of the critical drop, 
coming either from 
NVT-Seeding or from Umbrella Sampling simulations (see section \ref{sec:us}).  In these runs, the pressure is fixed to the virial value of the 
simulations were the critical clusters were previously equilibrated. 
According to Eq. \ref{eqf+}, the slope of Fig. \ref{f+} divided by 2 gives $f^+$. Multiplying such $f^+$ by the Zeldovich factor we get an 
estimate of the kinetic pre-factor, $A_0^{af}$, that does not rely on the kinetic theory of gases. 
$A_0^{af}$
is reported 
it in table \ref{table:sizes} 
for a couple of critical clusters generated with 
NVT-Seeding (systems IV and VI).
$A_0^{af}$ is very close to 
$A_0^{ktg}$. This agreement suggests the 
validity of the kinetic theory of gases to estimate 
the attachment rate and makes the theoretical framework 
that supports the Seeding technique quite powerful given 
that, since $A_0^{ktg}$ can be used, 
only $R_c$, $\Delta p$ and the density of both phases
are required to get accurate estimates of $J$ in a wide range
of orders of magnitude. Note in Fig. \ref{deltaG}(c) that 
Seeding (black dots) has access to $J$ values many orders of magnitude lower
than US (red dots). 

\begin{table}[]
\begin{ruledtabular}
\begin{tabular}{cccccccccccccccc}
Label &
$N_T$ &
$\langle V \rangle$ &
$p_v$ &
$\rho_v$  &
$\log_{10}(J)$\\
\hline
BF-1 & 4000 & 57145 & 0.035 &  0.0700 & -9.235 \\
BF-2 & 4000 & 53456 & 0.036 &  0.0748 & -8.287 \\
\end{tabular}
\end{ruledtabular}
\caption{Data corresponding to the brute force calculations.}
\label{table:spontaneous}
\end{table}

The green dots in Fig. \ref{deltaG}(c)
correspond to rate estimates obtained in brute force NPT
molecular simulation runs performed at high supersaturations
where condensation occurs spontaneously from an unseeded
vapor. 
In such cases the nucleation rate can be estimated 
as $J=1/(t <V>)$, where $<V>$ is the average volume 
before nucleation and $t$ is the nucleation time 
averaged over a number of independent trajectories (typically
20 in our case). $N_T$, $V$, the vapor pressure and density, and $J$
for the two states where we studied spontaneous condensation 
are reported in Table \ref{table:spontaneous}. 
In Fig. \ref{deltaG}(c) we show that $J$ estimates from Seeding and from spontaneous nucleation 
are consistent with each other, which further indicates
the ability of
Seeding to predict nucleation rates. It is worth mentioning here that NVT-Seeding and spontaneous nucleation are complementary techniques.
On the one hand, the former does not have access to such high supersaturations
given the difficulty to equilibrate small clusters in the NVT 
ensemble \cite{seedingNVT,montero2020interfacial}. 
That said, it would be nonsense using Seeding where nucleation 
occurs spontaneously in a straightforward manner. 
On the other hand, spontaneous nucleation is limited to a narrow
window of nucleation rates (that enabled by computational time)
whereas Seeding has access to extremely low rates. 

\textcolor{black}{We would like to end this section by discussing finite
size effects, which could be present if a nucleus sees its replica
through periodic boundary conditions. 
On the one hand, we made sure that the density of the outer phase reaches
a plateau before L/2 by looking at radial density profiles such as that shown
in Fig. \ref{rdp}. On the other hand, we note that the box side of system X is 
more than twice than those of the other systems. By looking at Figs. \ref{deltaG} (a), (b) and (c) one can see that the results from system X are fully consistent with those inferred
from the other systems, which strongly supports the absence of noticeable finite
size effects in our simulations.}

\begin{table}[]
\begin{ruledtabular}
\begin{tabular}{cccccccccccccccc}
Label &
$L$ &
$N_T$ &
$\Delta G_c/(k_BT)$ &
$A_{0}^{af}$  &
$\log_{10}(J)$\\
\hline
US-1 & 39.112 & 4000 & 17.7 &  0.041 &  -10.3 \\
US-2 & 38.501 & 4000 & 40.7 &  0.039 & -20.3 \\
\end{tabular}
\end{ruledtabular}
\caption{Data corresponding to the US calculations.}
\label{tablaUS}
\end{table}

\section{Umbrella Sampling}
\label{sec:us}

As previously indicated, to validate the Seeding results we used the US technique. We followed Refs. \cite{JCP_1998_109_09901,gonzalezPCCP2014} to compute
$\Delta G_c$ for two different vapor pressures: 
p=0.031 and p=0.033. Details on the simulation 
box size and number of particles in the systems
used to perform the US calculations are given 
in table \ref{tablaUS}.

The free energy associated to the formation of an 
$N$ particle cluster drop can be obtained from:
\begin{equation}
    \Delta G(N)=-k_BT \ln[P(N)],
    \label{freeenergy}
\end{equation}
where $P(N)$ is the probability distribution of $N$.
Our criterion to compute $N$ is described in Section 
\ref{sec:J}. It is important to note that 
even though different criteria may
give different $N$ for a given configuration, the
height of an US free energy barrier does not depend
on the criterion to determine the cluster size \cite{filion:244115}. Therefore, contrary to what
happens in Seeding, the US method does not depend
on the specific criterion to determine the nucleus size. 
This is why it is important to validate the Seeding
method with other tecniques such us US.

With conventional NpT simulations at the
selected pressures $P(N)$ can only be sampled up to $N \sim 40$ while the critical cluster is much larger in this regime. To sample the rest of the free energy barrier a biasing potential, $U_{bias}$, is added to the 
original hamiltonian:
\begin{equation}
    U_{bias}=\frac{1}{2}k_{bias}\left(N-N_0\right)^2,
    \label{biaspotential}
\end{equation}
where $N_0$ controls the cluster size around which the sampling will be centred and $k$ the width of such sampling. 
Tens of overlapping sampling ``windows'' centered at 
different $N_0$ values are required
to reconstruct the whole free energy barrier. 
The effect of the bias potential on the calculation 
of the free energy barrier is removed as follows \cite{torrie1974monte}:
\begin{equation}
\Delta G(N)=-k_B T\ln\left<\frac{\chi_N}{e^{-U_{bias}/(k_BT)}}\right>+C
    \label{USenergy}
\end{equation}
where $\chi_N$ is the fraction of clusters with $N$ particles that appear within a certain window and $C$ is a constant. The constant is
obtained by gluing together the first part of the energy barrier evaluated without the biasing potential (Eq. \ref{freeenergy} ) with the rest of the windows. The result is the whole free energy barrier.

To compute each window we use the hybrid Molecular Dynamics-Monte Carlo scheme labelled as HMC(nM-NpT)/US in Ref. \cite{gonzalezPCCP2014}. From the starting configuration, random velocities are assigned to every particle according to a Maxwell-Boltzmann distribution and a short ($\Delta t$ 19.2 Lennard Jones times) MD simulation is run  for generating a new configuration, which is accepted with probability  $\min[1,\exp[´- (U_{bias}(\Delta t)-U_{bias}(0))/(k_BT)]]$. Either 
in case of acceptance or rejection new random velocities are assigned at the beginning of each short MD cycle.  
For each window, 10000 of such cycles were performed for equlibration and 60000 to obtain the free energy barrier. We used $k_{bias}=0.04 k_BT$ in the biasing potential (Eq. \ref{biaspotential}), 
which gives an acceptance rate of $\sim 25\%$.

\begin{figure}[h!]
\includegraphics[width=0.9\linewidth]{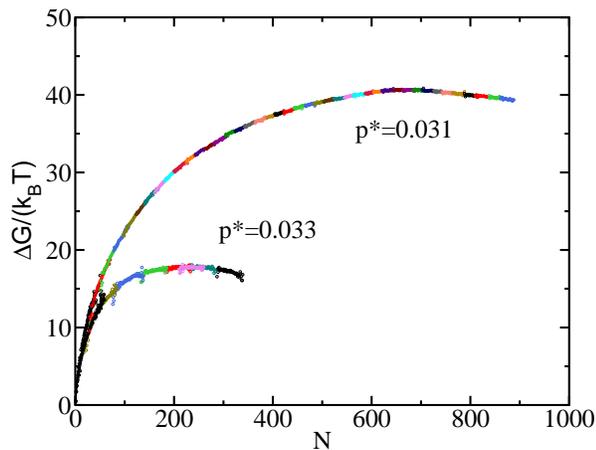}
\caption{Free energy for two different pressures (p=0.031 and p=0.033) versus the number of particles in the drop as obtained from US calculations. The different colours represent the different windows performed.}
\label{barrerasUS}
\end{figure}

In figure \ref{barrerasUS} we plot both free energy barriers, being $\Delta G_c=17.7 k_BT$ for p=0.033 and $\Delta G_c=40.7 k_BT$ for p=0.031 (also reported in 
table \ref{tablaUS}). As already discussed, 
the agreement between US and Seeding is excellent 
(see Fig. \ref{deltaG}(b)).

Additionally, we compute the kinetic pre-factor $A_0^{af}$ (Eq. \ref{eqA0})
to obtain the nucleation rate (Eq. \ref{eq:rate}).
To do that, we launch tens of unbiased trajectories from independent
configurations at the barrier top in order to 
compute the attachment rate via Eq. \ref{eqf+}.
The Zeldovich factor (Eq. \ref{zeq}) can be obtained
by numerically calculating the curvature
of $\Delta G(N)$ at the barrier top.
We report $A_0^{af}$ thus calculated and the corresponding
$J$ 
in table \ref{tablaUS}.
As previously discussed, $J$ from US is fully consistent
with that coming from Seeding (see Fig. \ref{deltaG}
(c)). 

In summary, we have compared Seeding, that relies
on the theoretical assumptions by CNT and $ktg$
and depends on the criterion employed to determine
the cluster size, with US, that does not have these
limitations. We have obtained an excellent agreement
between both techniques. 
This is very good news because Seeding is much
more efficient than US and has access to much lower
values of the nucleation rate.

\begin{table*}[]
\begin{ruledtabular}
\begin{tabular}{cccccccccccccccc}
Label &
$L$ &
$N_T$ &
$\rho_{vap} $ &
$\rho_{l}$  &
$p_{vap}$ &
$p_{l}$ &
$\Delta p$ &
$R_c$ &
$\gamma$ &
$\Delta G_c/(k_BT)$ &
$A_{0}^{BK}$  &
$\log_{10}(J)$\\

\hline
I   & 36.731 & 30795 & 0.03765 & 0.6453 & 0.02365 & -0.02601 & 0.0497 &  7.35 & 0.1826 &  52.7 & 0.341  & -23.5 \\
II  & 36.731 & 30342 & 0.03834 & 0.6484 & 0.02398 & -0.01914 & 0.0431 &  8.50 & 0.1832 & 70.6  & 0.342    &-31.3 \\
III & 36.731 & 29760 & 0.03875 & 0.6501 & 0.02419 & -0.01503 & 0.0392 &  9.53 & 0.1869 & 90.6 & 0.345    & -40.0 \\
IV  & 36.731 & 29034 & 0.03907 & 0.6514 & 0.02433 & -0.01191 & 0.0362 & 10.52 & 0.1906 & 112.5 & 0.348    &  -49.5  \\
V   & 36.731 & 28147 & 0.03949 & 0.6530 & 0.02453 & -0.00776 & 0.0323 & 11.50 & 0.1857 & 131.1 &  0.344    & -57.6 \\
VI  & 36.731 & 27082 & 0.03972 & 0.6539 & 0.02464 & -0.00558 & 0.0302 & 12.45 & 0.1881 & 155.7 &  0.346    & -68.3 \\
\end{tabular}
\end{ruledtabular}
\caption{NVT-Seeding data for the different bubbles studied in this work at T=0.785.}
\label{table:bubbles}
\end{table*}

\begin{figure*}[h!]
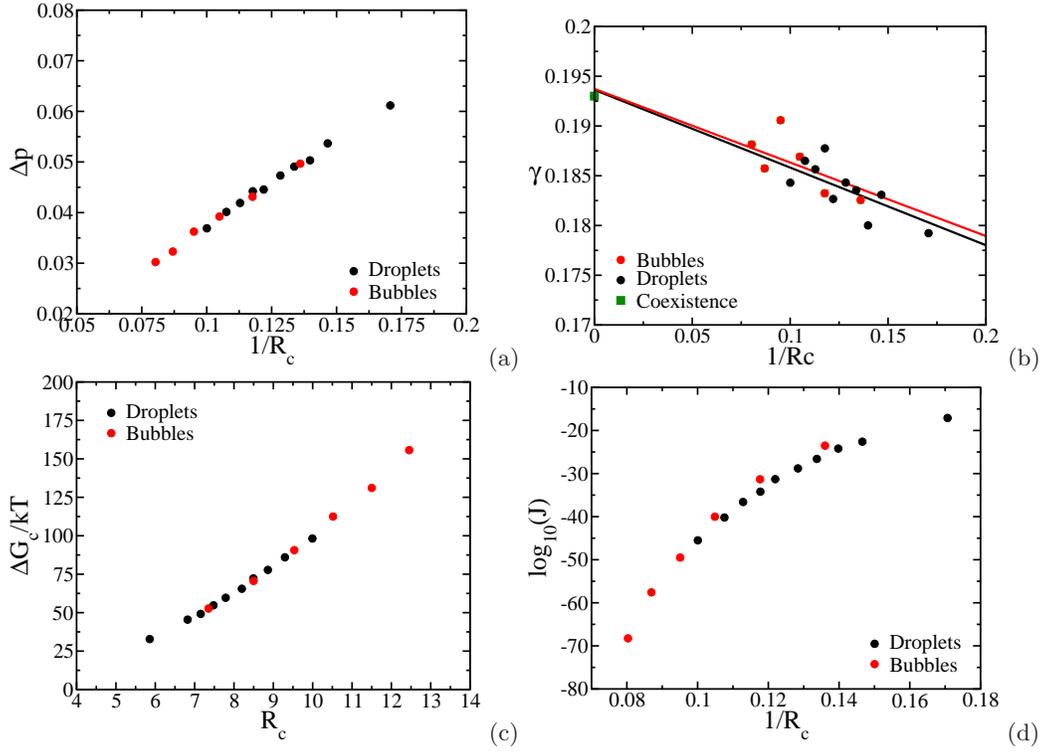

		\includegraphics[width=0.35\linewidth]{deltaP_invR.eps} (a) 
			\includegraphics[width=0.35\linewidth]{gamma_vs_1R.eps} (b)\\
		\includegraphics[width=0.35\linewidth]{dG_r.eps} (c)
			\includegraphics[width=0.35\linewidth]{logJ_invR.eps} (d)
	\caption{(a) $\Delta p$ vs. $1/R_c$, (b) $\gamma$ vs. $1/R_c$, (c) $\Delta G_c$ vs. $R_c$, and (d) $log_{10}J$ vs. $1/R_c$ for droplets (black symbols) and bubbles (red symbols) as obtained from NVT-Seeding.	}
\label{gammavs1R}
\end{figure*}

\section{Condensation vs boiling}

\subsection{Comparison for a given $R_c$}

We have studied quite recently the 
nucleation of bubbles for the same Lennard Jones model
employed here \cite{seedingNVT}. 
Since the study was performed at the same temperature,
the question that naturally arises is whether bubbles and drops with the same
radius have the same interfacial properties. 
To establish the comparison we have repeated the analysis performed
in Ref. \cite{seedingNVT} 
because in such work we used 
0.026 as the coexistence pressure instead of 0.0267. 
We took the 0.026 value from a paper published more than 
a decade ago \cite{wang2008homogeneous}. 
However, we have recomputed more carefully the coexistence pressure 
at $T=0.785$ and obtain $p=0.0267$ instead, which is the value
we use in this work. The difference is subtle, but given 
that the pressure inside the nucleating phase is obtained
by integrating from the coexistence pressure (see Eq. \ref{eq:intterm})
it is very important to use an accurate value for the latter. 

\begin{figure}[h!]
	\includegraphics[width=0.85\linewidth]{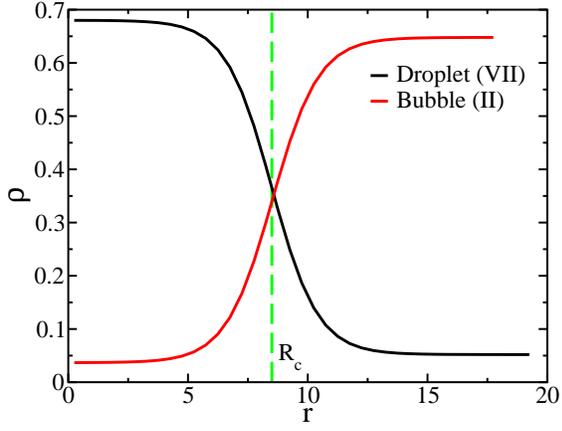}
	\caption{\textcolor{black}{Radial density profiles of drop VII and bubble II, compared. They have almost identical radius, $R_c$.}}
\label{fig:dpc}
\end{figure}

The simulation data for different bubbles equilibrated at T=0.785
in the NVT ensemble are reported in table \ref{table:bubbles}.
The values of $R_c$ corresponding to each system (obtained
with the equi-density criterion as discussed in section \ref{sec:rc} and in Ref. \cite{seedingNVT}) are very
close to those recently reported by ourselves \cite{seedingNVT}.
However, the values of $\Delta p$ here reported are not identical
to those of Ref.\cite{seedingNVT} due to the coexistence pressure
issue discussed above. In Fig.\ref{gammavs1R}(a) we plot
$\Delta p$ vs $1/R_c$
for bubbles and and drops at T=0.785.
Drops and bubbles of the same size 
have the same $\Delta p$, which is perhaps the most important result
of the paper. Note that, for a given $R_c$, the 
pressures of the external and the internal phases are
not the same if one compares cavitation and condensation. What is the same is the pressure difference between the internal and the 
external phases. For instance, let's focus on the 
case of drop VII and bubble II, both with $R_c \approx 8.5$. 
\textcolor{black}{In Fig. \ref{fig:dpc} we compare their radial
density profiles. The density of the liquid inside the drop is different from 
that of the liquid outside the bubble. Also, the density of the vapor inside the bubble
is different from that of the vapor outside the drop.}
The bubble is surrounded by a liquid of pressure
-0.01914 whereas the drop by a vapor
of pressure 0.0301: the pressures of the external 
phases do not even have the same sign. 
The bubble and the drop also have very different pressures:
0.02398 and 0.07432 respectively.
Despite the fact that
the external and the internal pressures are very different, $\Delta p$ is not: 0.043 and 0.044 for the bubble and the drop 
respectively.

According to the Laplace equation, that $\Delta p(R_c)$ is the same
for drops and bubbles, implies that $\gamma$ 
must be the also same 
regardless the identity of the internal and the external 
phases.
In Fig. \ref{gammavs1R} (b) we plot $\gamma$ vs
$1/R_c$ for bubbles and drops and find that, indeed, 
they have the same $\gamma$ within our statistical noise.
Of course, attending to Eq. \ref{eq:deltag}, 
$\Delta G_c$, that only depends on $R_c$ and $\Delta p$, is also the same for a given $R_c$, as illustrated in Fig. \ref{gammavs1R}(c).

The nucleation rate for bubbles with a given
$R_c$  is close to the corresponding
drop, but is not exactly the same, given that
the kinetic pre-factor is not identical. 
In the case of bubble nucleation we have carefully 
assessed \cite{seedingNpT} that the following expression by Blander and Katz (BK) provides a good approximation for $A_0$:
\begin{equation}
    A_0^{BK}=\sqrt{\frac{\Delta p R_c}{\pi m}},
\end{equation}
which is very similar, but with a missing
$(\rho_{vap}/\rho_{l})$ factor with respect to the $ktg$ expression
we use for drop condensation (Eq. \ref{eq:a0ktg}).
The $A_0^{BK}$
values we use in our Seeding predictions
of bubble cavitation are reported in 
table \ref{table:bubbles} alongside 
the resulting values of $J$ obtained 
as $J=\rho_{l} A_0^{BK} \exp[{-\Delta G_c/(k_BT)}$].
As it can be seen in Fig. \ref{gammavs1R}(d), 
the rate for bubbles and drops for a given
$R_c$ is quite similar, although it is systematically lower for the latter
due to the $\rho_{vap}/\rho_{l}$ factor
previously mentioned. 

Condensation and cavitation have already been compared in 
the literature  \cite{guermeur1985density,schmelzer2019entropy,macdowell2006nucleation,JCP_2004_120_05293,block2010curvature,binder2012beyond,caupin2015}.
However, there are only few cases in which $\gamma$
has been compared for a given temperature as a
function of the droplet/bubble size \cite{block2010curvature,binder2012beyond} as we
do in this work. Establishing such comparison 
in experiments is difficult because it is not
possible to detect the critical nucleus. 
In simulations the nucleus can be visualized, but
computing $\gamma$ is a hard task. 
It requires either computing the free energy 
of a system with the nucleus inside  \cite{blokhuis2006thermodynamic,binder2012beyond}
or, more easily, computing the nucleus size and
using a theory to infer $\gamma$ \cite{baiJCP2006,carignano} as we do in this work. 
In either approach, one has to deal with the
arbitrariness
of establishing a location for the interface. 

In our case, we found in a recent work 
by ``trial and error'' 
that the equi-density surface  
gives good results for cavitation \cite{seedingNpT}.
By ``good results'' we mean that Seeding predictions 
of nucleation are consistent with those coming from independent
methods that do not rely on a precise definition 
of the nucleus size. 
In this work we have demonstrated that 
the same criterion to locate the interface
is successful in condensation. 
Therefore, one of our main findings is that the equi-density surface is the one that provides
good predictions when CNT is used both for cavitation and for 
condensation. 
This means that the equi-density surface
can be identified with the surface of tension, 
which is the one for which CNT works and the
Laplace equation holds (see section \ref{cnt})
\cite{kashchievbook,montero2020interfacial,troster2012numerical,statt2015finite} .
We believe that identifying the surface
of tension with the equi-density surface 
both for cavitation and condensation is
an important finding of our work. 
This leads to the relevant conclusion that 
condensation and cavitation are two sides of the same coin
in the sense that they share the same surface tension.

In Ref. \cite{block2010curvature} 
$\gamma$ was found to be quite different for both phenomena, 
but  the comparison was not established for the surface
of tension but for the equimolar surface. 
In Ref. \cite{binder2012beyond}, however, the comparison 
was established for the first time for 
the surface of tension and, although
$\gamma$ was similar for condensation and cavitation, there
were significant differences that need to be further investigated
in order to match our work with that of Ref. \cite{binder2012beyond}.

\subsection{Comparison for a given metastability degree}
\begin{figure}[h!]
	\includegraphics[width=0.85\linewidth]{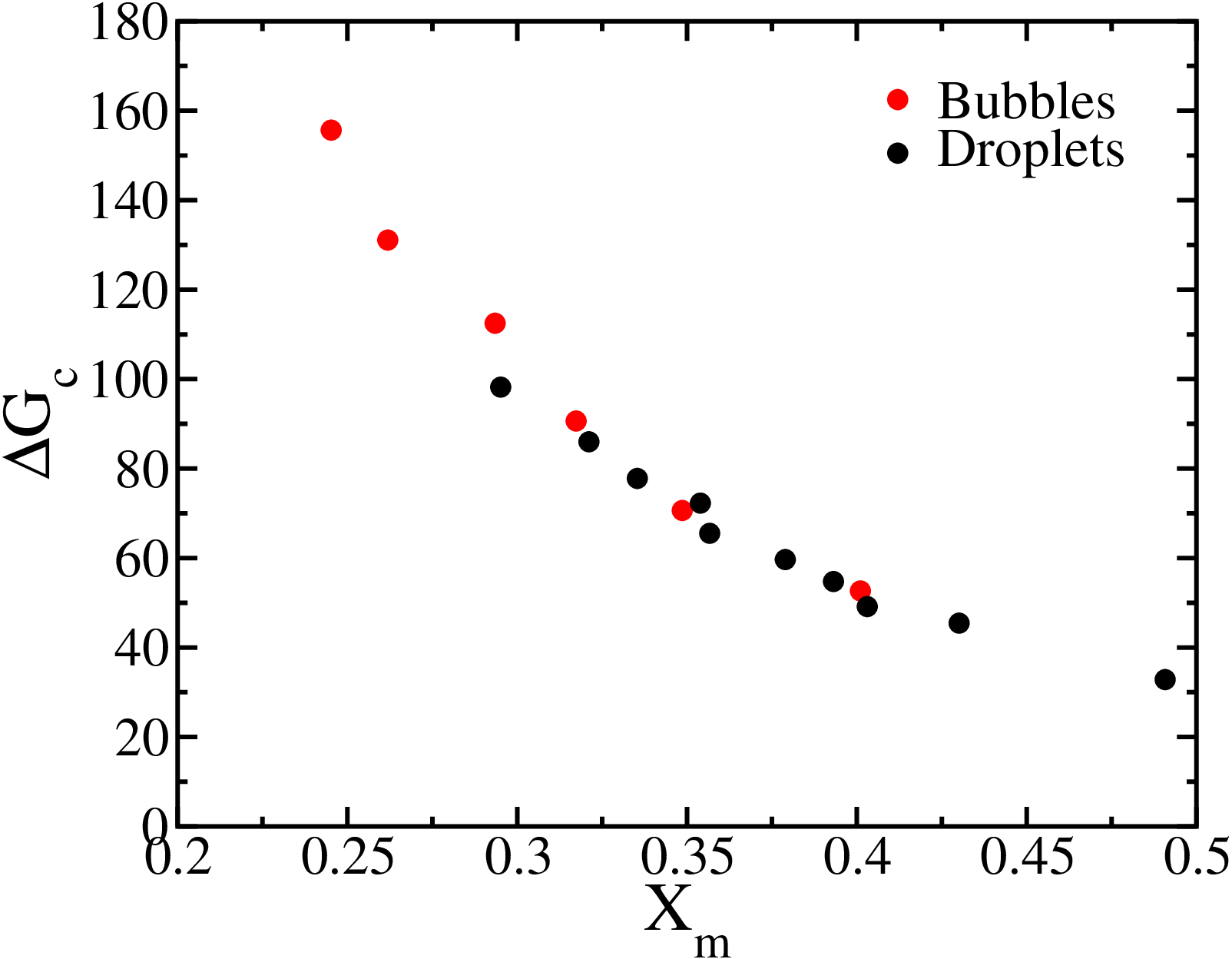}
	\caption{\textcolor{black}{Nucleation free energy barrier for drops and bubbles (see legend) as a function of the metastability degree, $X_m$, defined in 
	Eq. \ref{eq:X_m}}.}
\label{fig:deltaspin}
\end{figure}

\textcolor{black}{
In Ref. \cite{shen2001density} it was proposed in a DFT study that the work of formation of critical bubbles 
studied at different temperatures collapse when plotted  
against the metastability degree, $X_m$, quantified as:
\begin{equation}
    X_m = \frac{\mu_{nuc}-\mu_{coex}}{\mu_{spinodal}-\mu_{coex}}
\label{eq:X_m}    
\end{equation}
} \textcolor{black}{
where $\mu_{nuc}$ is the chemical potential of the parent phase at the conditions
where 
nucleation is studied, $\mu_{coex}$ is the coexistence chemical potential at 
the same temperature and at coexistence pressure, and $\mu_{spinodal}$ is the chemical potential at the same temperature 
but at the pressure where spinodal decomposition takes place.
To estimate the spinodal pressure we 
run NpT simulations of the bulk liquid and vapor phases with
4000 particles. We estimate the spinodal decomposition
pressure as that for which we the system undergoes a phase transition without
any induction period, right after the start of the 
simulation. 
Both chemical potential differences in Eq. \ref{eq:X_m}  can be easily obtained by 
numerically integrating the molar volume along pressure at constant temperature. 
The denominator is the maximum possible metastability whereas
the numerator is the actual metastability of the state where
nucleation is studied. Therefore, $X_m$ varies from 0
at coexistence, to 1 at spinodal decomposition. 
The mestastability degree above described can be computed for drop 
as well as for bubble nucleation. Therefore, we have the chance
to compare nucleation free energy barriers for drops and bubbles 
as a function of $X_m$. The comparison, shown in Fig. \ref{fig:deltaspin}, 
reveals the interesting conclusion that $\Delta G_c$ for bubble and
drop nucleation is the same for
a given metastbility degree.  Therefore, not only nucleation barriers
at different temperatures can be collapsed via the metastability degree as
proposed in Ref. \cite{shen2001density},
but also bubble and drop nucleation data match for a given metastability degree.}

\subsection{Tolman Legth}

\begin{figure}[h!]
		\includegraphics[width=0.85\linewidth]{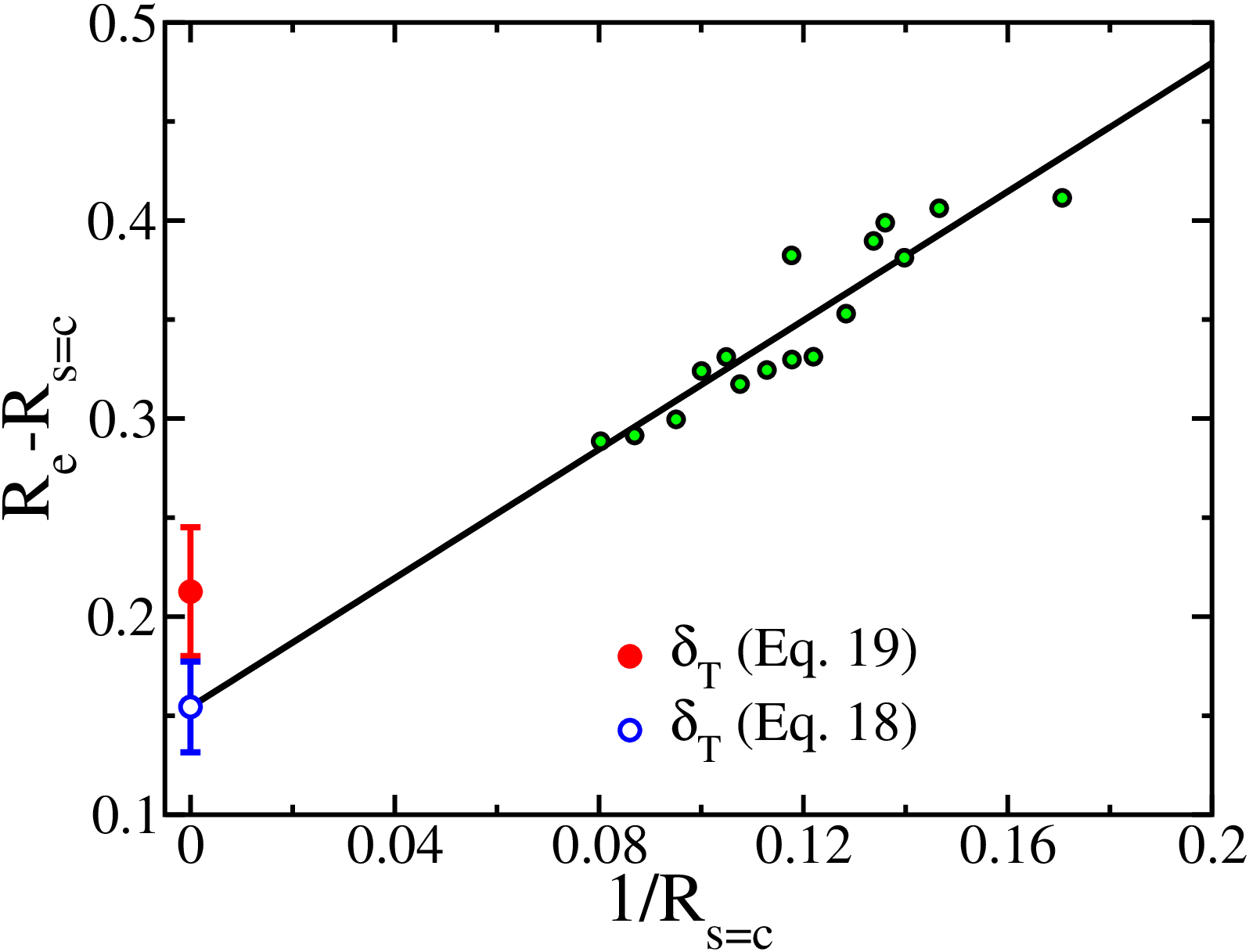}(a)\\	
	\includegraphics[width=0.85\linewidth]{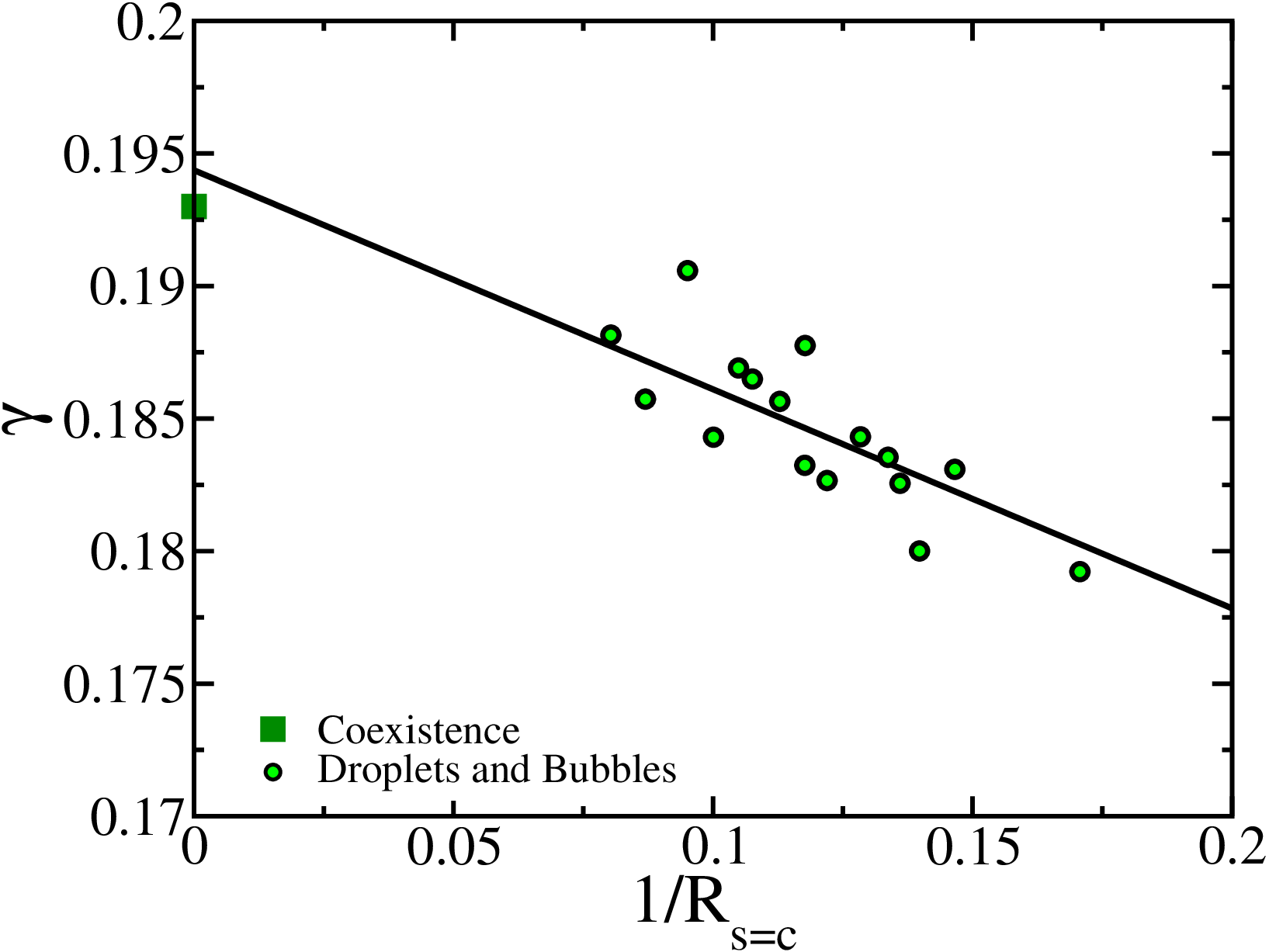}(b)\\
	\caption{(a) $R_e-R_{s=c}$ and (b) $\gamma$ vs. $1/R_{s=c}$ for drops and
	bubbles together.
	}
\label{fig:tolman}
\end{figure}

Since bubbles and drops of the same radius
have the same interfacial properties, we can
use the data coming from both systems altogether 
in order to compute the Tolmann Lenght, $\delta_T$, 
which is defined as \cite{tolman1949effect,blokhuis2006thermodynamic}:
\begin{equation}
	\delta_{Tolman}= \lim_{R_s \rightarrow \infty} (R_e - R_s)
	\label{eq:tolmanrig}
\end{equation}
where $R_e$ is the Gibbs equi-molar radius and $R_s$ is the radius
of the surface of tension. We identify $R_s$ with $R_c$ (the equi-density radius)
given that (i) we obtain good predictions of nucleation when we use $R_c$ and
(ii)  $R_s$ is the radius that enters CNT \cite{kashchievbook,montero2020interfacial,troster2012numerical,statt2015finite}. To underline the 
fact that we identify $R_c$ with $R_s$ we 
label $R_c$ as $R_{s=c}$ in the following figures. 
$R_e$ can be easily computed from the radial density profiles \cite{seedingNpT,montero2020interfacial}.
In Fig. \ref{fig:tolman}(a) we show $R_e-R_{s=c}$ versus 1/$R_{s=c}$ for all data (either bubbles
or drops) coming from this work. The extrapolation to $1/R_{s=c} =0$ provides an estimate
of $\delta_{Tolman}$, indicated with an empty blue dot in the figure. We obtain $\delta_{Tolman}=0.15 \pm 0.02$.
We showed in a recent paper, in which we analysed spherical hard sphere crystals in equilibrium with the fluid, that $\delta_{Tolman}$ can be also estimated by fitting 
$\gamma$ to the following expression:
\begin{equation}
   \gamma = \gamma_{0} \left(1 - 2 \frac{\delta_{T}}{R_s}\right),
\label{eq:tolmanfit}
\end{equation}
where $\gamma_0$ is the value of $\gamma$ at coexistence at the temperature of interest and $\delta_T$ is the 
fitting parameter that serves as an estimate for
$\delta_{Tolman}$ \cite{montero2020interfacial}. 
\textcolor{black}{This approach is similar in spirit to those that include
$\gamma$ given by Eq. \ref{eq:tolmanfit} in CNT to fit free energy 
barriers obtained by rare event methods \cite{gallo2020nucleation,menzl2016molecular}.}
Again, we identify here $R_s$ with $R_c$.  Consequently, we use the $\gamma$ data 
coming from such radius (that reported
in tables \ref{table:sizes} and \ref{table:bubbles}) to obtain an estimate of $\delta_T$ with the expression above. 
The data of $\gamma$ vs $1/R_{s=c}$ are shown in green
in Fig. \ref{fig:tolman}(b). The solid line is a linear fit of $\gamma$ vs $1/R_{s=c}$ 
which includes $\gamma_0$ (the green square in the figure). 
The $\delta_T$ value coming from such fit, $\delta_T = 0.21 \pm 0.03$, is shown with a red dot in Fig. \ref{fig:tolman}(a). 
Both values, $\delta_{Tolman}$ obtained via Eq. \ref{eq:tolmanrig} (blue dot in Fig. \ref{fig:tolman}(a))
and $\delta_T$ coming from Eq. \ref{eq:tolmanfit} (red dot the same figure), are consistent with each
other within the statistical uncertainty of our estimates. 
This corroborates the idea, 
recently checked for the first time for 
hard sphere crystals \cite{montero2020interfacial}, that the Tolman lenght can be obtained either from Eq. \ref{eq:tolmanfit} or
from Eq. \ref{eq:tolmanrig}. Hence, this idea seems to be a general one pertaining not only to the 
crystal-fuid equilibrium but also to the liquid-vapor one. 

This study may shed some light in the intense literature
debate about the magnitude and sign of the Tolman 
length \cite{blokhuis2006thermodynamic,block2010curvature,sampayo,malijevsky2012perspective,binder2012beyond,size-dependendentgamma2005,ceriottitolman2018,JCP_1984_81_00530,lau2015surface,vrabec2006comprehensive,wilhelmsen2015tolman,joswiak2016energetic,schmelzer2019entropy,richard2018crystallization,caupinPRL2016curvature}. We 
obtain a Tolman length of about twenty per cent 
the particle diameter. Its sign is positive, 
which means that $\gamma$ decreases when one
moves away from coexistence at constant temperature. 

\section{Conclusions}

The main conclusions we draw from our work are the 
following:
\begin{enumerate}
    \item We have used NVT-Seeding to investigate droplet nucleation in a supersaturated Lennard-Jones vapor. The results obtained from this technique are consistent with: (i) independent calculations of the nucleation free energy barrier performed with Umbrella Sampling (ii) the surface tension of a flat interface obtained from the pressure tensor in a vapor-liquid coexistence simulation (iii) the drop nucleation rate obtained both with US and in brute force spontaneous nucleation simulations. 
    \item NVT-Seeding requires defining the radius of a droplet equilibrated in the NVT ensemble. The radius definition that passes the consistency tests mentioned in the previous paragraph is that given by the surface where the density is average between that of the interior and that of the exterior phases. Such radius definition was also successful in our earlier studies of bubble nucleation \cite{seedingNpT,seedingNVT}. Therefore, we identify this ``equi-density'' radius with the radius of tension, $R_s$. 
    \item The good performance of Seeding strongly supports the use of CNT to describe nucleation.
    However, the capillarity approximation (that $\gamma$ is curvature independent) does not provide good results.
    A $\gamma$ dependent on the curvature of the critical nucleus must be plugged into the the theory. Therefore, the theory, although powerful, requires the involvement of simulations given that the  $\gamma$-curvature dependence is obtained by computing the size of the critical cluster at different pressures. 
    \item The kinetic theory of gases provides very good estimates of the kinetic pre-factor of the condensation nucleation rate. This makes the theoretical framework very powerful given that only 
    the size of the critical cluster, the density of the external phase and the bulk phases equations of state are needed to estimate nucleation rates.
    \item We compare NVT-Seeding results of droplets with those obtained for bubbles and find that, for a given temperature, bubbles and droplets of the same radius have, within the accuracy of our method, the same pressure difference with the surrounding medium. Therefore, bubbles and droplets of the same size have the same surface tension and the same nucleation free energy barrier. In this respect, 
    condensation and boiling can be seen as two 
    sides of the same coin. Such duality is only 
    verified if the size of the critical nucleus (either a bubble or a drop) is determined with the equi-density radius (our empirical definition of
    the surface of tension).
    \item We estimate the Tolman  length, $\delta_T$, by extrapolating to infinite-size drops/bubbles the difference between the equimolar radius, $R_e$, and $R_s$. Such $\delta_T$ is consistent with that obtained by linearly fitting $\gamma(1/R_c)$, in accordance with our recent work of hard sphere crystals \cite{montero2020interfacial}.
\end{enumerate}

\section{Acknowledgments}
This work was funded by grants FIS2016/78117-P
and PID2019-105898GB-C21 of the MEC.
The authors acknowledge the computer resources and technical assistance provided by the RES. 
P. R. thanks a doctoral grant from UCM.
P. M. H. acknowledges financial support from the FPI grant no. BES- 671
2017-080074. We also thank the reviewers of this work
for their suggestions. 

\clearpage

\bibliographystyle{ieeetr}

\end{document}